\documentclass[preprint, showpacs, preprintnumbers, amsmath,amssymb]{revtex4}
\usepackage{graphicx}
\usepackage{dcolumn}
\usepackage{bm}
\usepackage{epsfig}

\begin{document}

\title{Range-based attacks on links in random scale-free networks}

\author{Baihua Gong$^{1}$}
\author{Jun Liu$^{1}$}
\author{Liang Huang$^{1}$}
\author{Kongqing Yang$^{1,2}$}
\author{Lei Yang$^{3,4}$\footnote{Corresponding author}}
\affiliation{\\
$^{1}$Institute of Theoretical Physics, Lanzhou University, Lanzhou 730000, China\\
$^{2}$Institute of Applied Physics, Jimei University, Xiamen 361021, China\\
$^{3}$Institute of Modern Physics, Chinese Academy of Science,  Lanzhou 730000, China\\
$^{4}$Physics department, Beijing-Hong Kong-Singapore Joint Centre for Nonlinear and Complex Systems
(Hongkong), Hong Kong Baptist University, Hong Kong, China}

\date{\today}

\begin{abstract}
$Range$ and $load$ play keys on the problem of attacking on links in random scale-free (RSF) networks. In
this Brief Report we obtain the relation between $range$ and $load$ in RSF networks analytically by the
generating function theory, and then give an estimation about the impact of attacks on the $efficiency$ of
the network. The analytical results show that short range attacks are more destructive for RSF networks, and
are confirmed numerically. Further our results are consistent with the former literature (Physical Review E
\textbf{66}, 065103(R) (2002)).
\end{abstract}

 \maketitle

Attacking on complex networks, especially in the context of the Internet and biological networks, has been an
interesting issue, and different aspects of attacking have been analyzed recently \cite{attack}.  Many works
focus on attacks on nodes, and the strategies provided include random attacks, degree-based attacks, etc.
\cite{node}. Also, some works consider attacks on links, and the strategies include the range-based attacks,
load-based attacks, etc. \cite{link}.

Adison E. Motter, Takashi Nishikawa and Ying-Cheng Lai \cite{laiyc} studied attacks on links in scale-free
networks basing on $range$. $Range$ is introduced by Watts \cite{range}  to characterize different types of
links in networks: the range of a link $l_{ij}$ connecting nodes $i$ and $j$ is defined as the length of the
shortest path between the nodes $i$ and $j$ in the absence of  $l_{ij}$. The small-world model introduced by
Watts and Strogatz \cite{ws} (WS model) is more sensitive to attacks on long $range$ links connecting nodes
that would otherwise be separated by a long distance. It is not true for many scale-free networks, though
most of them also have a short average path length like the WS model. Adison E. Motter, Takashi Nishikawa and
Ying-Cheng Lai found that short-range links other than long-range ones are vital for efficient communication
between nodes in these networks. They argued that the average shortest path is a global quantity which is
mainly determined by links with large $load$, where the $load$ of a link is defined as the number of shortest
paths passing through this link \cite{load1,load2}. And for scale-free networks, with exponent in a finite
interval around $3$, due to the heterogeneous  degree distribution, the $load$ is on average larger for links
with shorter $range$, making the short-range attacks more destructive.

In this brief report, employing the generating function theory, we first derive an approximate relation
between $R(k_1,k_2)$ and $L(k_1,k_2)$ for RSF networks analytically, where $R(k_1,k_2)$ and  $L(k_1,k_2)$ are
defined as the expected value of $range$ and $load$ respectively for links between nodes with given degree
$k_1$ and $k_2$. We then give an estimation about the decrement of $efficiency$ as a function of $R(k_1,k_2)$
and $L(k_1,k_2)$, showing that short-range attacks are more destructive for RSF networks. Numerical
simulations are also performed, confirming our analytical results.

To study ranged-based attacks on links in RSF networks, we measure the $efficiency$ of the network as each
link is removed. The $efficiency$ of a network with size $N$ is defined as \cite{efficiency}:
\begin{equation}\label{edef}
E=\frac{2}{N(N-1)}\sum \frac{1}{d_{ij}},
\end{equation}
where $d_{ij}$ denotes the length of the shortest path between the node-pair $(i,j)$, the sum is over all
pairs of nodes in the network. The $efficiency$ defined above has a finite value even for disconnected
networks, and larger values of $E$ correspond to more efficient networks.

When a link is removed from the network, the $efficiency$ of the network generally decreases. The decrement
of $efficiency$ involves two quantities: (1) the number of node-pairs whose geodesic lengths increase; (2)
the average increment of the geodesic lengths of these node-pairs. The first quantity is related to the
$load$ of the removed link, and the second quantity is related immediately to the $range$ of the removed
link.

For RSF networks, the expected value of the geodesic length of node-pairs with given degree $k_{1}$ and
$k_{2}$ is
\begin{equation*}
d(k_{1,}k_{2})=\sum_{i=1}ip^{i}\left( k_{1,}k_{2}\right),
\end{equation*}
where $p^{i}(k_{1,}k_{2})$ is the probability that the node-pair with given degree $k_{1}$ and $k_{2}$ has a geodesic length $i$.
For RSF networks, we have \cite{pexp}
\begin{equation}\label{pexp}
p^{1}(k_{1,}k_{2})\approx \frac{k_1 k_2}{2N z_1},
\end{equation}
where $N$ is the number of nodes in the network, $z_1$ is the average number of first neighbors. By the
generating function formalism, we can obtain \cite{generating},
\begin{equation}\label{dexp}
d(k_{1,}k_{2})\approx 1+\frac{\ln (N\cdot z_1/(k_{1}\cdot k_{2}))}{\ln(z_{2}/z_{1})},
\end{equation}
where $z_{2}$ is the the average number of second neighbors. Accordingly the expected diameter of RSF
networks is \cite{generating}
\begin{equation}\label{edexp}
D \approx 1+\frac{\ln (N / z_{1})}{\ln(z_{2}/z_{1})}.
\end{equation}

Since the RSF network is totally random in all aspects other than the degree distribution, $R(k_1,k_2)$ is
thus equal to the expected value of geodesic length of nonadjacent node-pairs  with given degree $k_1-1$ and
$k_2-1$, that is
\begin{equation*}
R(k_{1,}k_{2})=\sum_{i=2} i\frac{p^{i}\left( k_{1}-1,k_{2}-1\right) }{1-p^{1}(k_{1}-1,k_{2}-1)},
\end{equation*}%
i.e.,
\begin{equation}\label{range1}
R(k_{1,}k_{2})=\frac{d(k_{1}-1,k_{2}-1)-p^{1}(k_{1}-1,k_{2}-1)}{1-p^{1}(k_{1}-1,k_{2}-1)}.
\end{equation}%
Combining Eq.~(\ref{pexp}), (\ref{dexp}) and (\ref{range1}), we can obtain
\begin{equation}\label{range2}
R(k_{1},k_{2})\approx 1+\frac {\frac{\ln (N z_{1}/((k_{1}-1)(k_{2}-1)))}{\ln
(z_{2}/z_{1})}}{1-\frac{(k_1-1)(k_2-1)}{2Nz_1}}.
\end{equation}

Furthermore, we assume that the network is spare, and can be seen as a tree with expected diameter $D$.
Consider a link $l_{ij}$ connecting node $i$ and $j$, where $i$ has a degree $k_1$ and $j$ has a degree
$k_2$. When removing $l_{ij}$, the network can be regarded as a tree $T_i$ rooted as $i$ or $T_j$ rooted as
$j$, both of which have a depth of $D-1$. Staring from the root $i$, the first layer has $k_1-1$ nodes, the
second layer has $z_1(k_1-1)$, and the $m$th ($0<m<D$) layer has $z_1^{m-1}(k_1-1)$ nodes. Similarly, the
$m$th layer of $T_j$ has $z_1^{m-1}(k_2-1)$ nodes. The geodesic path from nodes in the $d_1$th ($d_1<D-1$)
layer in $T_i$ to nodes in the $d_2$th ($d_2<=D-1-d_1$) layer in $T_j$ in are expected to pass $l_{ij}$,
which has contribution of $1$ to the load of $l_{ij}$. Thus the expected value of the load of $l_{ij}$ is
\begin{eqnarray}
L(k_1,k_2)=&&\sum_{d=1}^{D-2}{((k_1-1)\frac{(z_1-1)^d-1}{z_1-2}+1)(k_2-1)(z_1-1)^{D-d-2}}\nonumber\\
&&+((k_1-1)\frac{(z_1-1)^{D-1}-1}{z_1-2}+1)+(k_2-1)(z_1-1)^{D-2}\nonumber\\
=&&(\frac{(D-2)(z_1-1)^{D-2}}{z_1-2}-\frac{(z_1-1)^{D-2}-1}{(z_1-2)^2})(k_1-1)(k_2-1))\nonumber\\
&&+(k_1+k_2-2)\frac{(z_1-1)^{D-1}-1}{z_1-2}+1.
\end{eqnarray}
When $k_1>>z_1,k_2>>z_1$, the above equation can be rewritten as
\begin{equation}\label{lexp1}
L(k_1,k_2)=\frac{((D-2)(z_1-2)-1)(z_1-1)^{D-2}+1}{(z_1-2)^2}(k_1-1)(k_2-1),
\end{equation}
showing that $load$ is directly proportional to the product of $(k_{1}-1)$ and $(k_{2}-1)$ when $k_1$ and
$k_2$ are large enough. For simplicity, we rewrite Eq.~(\ref{lexp1}) as
\begin{equation}\label{lexp2}
L(k_1,k_2)=c (k_1-1)(k_2-1),
\end{equation}
where $c$ is the coefficient $\frac{((D-2)(z_1-2)-1)(z_1-1)^{D-2}+1}{(z_1-2)^2}$.

The above analytical results can be numerically verified in the
following. In our example, we take $N=10000, \lambda=3.5, k_0=6$.
We plot $R(k_1,k_2)$ in Fig.1, and $L(k_1,k_2)$ in the inset of
Fig.1. From the inset of Fig.1, it can be seen that $load$ is
directly proportional to the product $(k_{1}-1)(k_{2}-1)$ when
$(k_1-1)(k_2-1)$ are large enough.

%\begin{figure}[tbp]
%\centering \epsfig{file=RLKS.eps,width=9cm, height=7cm} \caption{Average $range$ as a function of the product
%$(k_{1}-1)(k_{2}-1)$ in RSF networks with $N=10^{4}, \lambda=3.5, m_0=6$, and the maximal degree $m_{\max
%}=500$. The solid line is  the theoretical curve and the hollow squares are simulation results. Inset:
%Average $load$ as a function of the product $(k_{1}-1)(k_{2}-1)$. Numerical data are obtained from $100$
%realizations. } \label{RLK}
%\end{figure}

%\begin{figure}
%\begin{center}
%\includegraphics [width=9cm]{RLKS.eps}
%\end{center}
% \caption{\label{RLK}
% Average $range$ as a function of the product
%$(k_{1}-1)(k_{2}-1)$ in RSF networks with $N=10^{4}, \lambda=3.5,
%m_0=6$, and the maximal degree $m_{\max }=500$. The solid line is
%the theoretical curve and the hollow squares are simulation
%results. Inset: Average $load$ (represented by $B$) as a function
%of the product $(k_{1}-1)(k_{2}-1)$. Numerical data are obtained
%from $100$ realizations.}
%\end{figure}

Combining Eq.~(\ref{range2}) and (\ref{lexp2}), we can see the $load$ and the $range$ have a negative
correlation, that is
\begin{equation}\label{range3}
R(L)\approx 1+\frac{\frac{\ln (cN\cdot z_{1}/L)} {\ln(z_{2}/z_{1})}}{1-\frac{L}{2Ncz_1}}.
\end{equation}%
This expression gives an estimation of the relation between $R$
and $L$, suggesting that short range links are expected to be
passed through a large number of shortest paths. Numerical
verification is presented in Fig.2, where theoretical estimation
and numerical results are plotted. When a link $l_{ij}$ is removed
from the network, the decrement of the $efficiency$ of the network
is approximately:
\begin{equation}
\Delta E\approx \frac{2}{N(N-1)}\sum_{(m,n)\in{\Gamma}}{ \frac{R(i,j)-1}{d_{mn}^2}}\approx
\frac{2(R(i,j)-1)L(i,j)}{D^2N(N-1)},
\end{equation}
where the set $\Gamma$ is all the node-pairs whose shortest length should increase as a result of the removal
of link $l_{ij}$, $R(i,j)$ and $L(i,j)$ are the $range$ and $load$ respectively of $l_{ij}$. Thus the product
$(R-1)L$  is a natural quantity to characterize the impact of removing a link on the $efficiency$. For RSF
networks,
\begin{equation}
\Delta E\approx h L(i,j)\frac{\frac{\ln (cN\cdot z_{1}/L(i,j))}{\ln (z_{2}/z_{1})}}{1-\frac{L}{2Ncz_1}},
\end{equation}
where $h$ is $\frac{2}{D^2N(N-1)}$. $\Delta E$ is an increasing function of $L$, thus a decreasing function
of $R$. It can be concluded that links with small range are more important for the efficiency of RSF
networks.

In summary, by investigating the expected $range$ and $load$ of links in RSF networks, we obtain an
approximate analytical relation between $range$ and $load$, and then give an estimation of the impact of
removal of links on the $efficiency$. Thus we prove analytically that attacks on short-range links are more
destructive for RSF networks. An insufficiency in our work is that $R(L)$  has a significant error comparing
with numerical results.  $R(L)$ given in Eq.~(\ref{range3}) is highly sensitive to some quantities such as
$c$ and $z_2$. These quantities are not obtained accurately due to the approximations used in our analysis,
such as the tree-like approximation necessary for the generating function method. Anyhow, the analytical
results in this brief report give a reasonable description for the trend of the true relation between $R$ and
$L$ for RSF networks.

The work is supported by National Hi-Tech Research and Development Program with Grant No.~2006AA09A102-08 and
National Basic Research Program of China with Grant No.~2007CB209603.  Lei Yang thanks the 100 Person Project
of the Chinese Academy of Sciences, the Hong Kong Research Grants Council (RGC) and the Hong Kong Baptist
University Faculty Research Grant (FRG) for their support.

\newpage

Fig.1 Average $range$ as a function of the product
$(k_{1}-1)(k_{2}-1)$ in RSF networks with $N=10^{4}, \lambda=3.5,
m_0=6$, and the maximal degree $m_{\max }=500$. The solid line is
the theoretical curve and the hollow squares are simulation
results. Inset: Average $load$ (represented by $B$) as a function
of the product $(k_{1}-1)(k_{2}-1)$. Numerical data are obtained
from $100$ realizations.\\

Fig.2 Average $load$ as a function of $range$. Square: theoretical
value; circle: averaged simulation value over 100 realizations;
(All the parameters are the same as in Fig.1) cross: simulation
value for $\lambda=3.0, N = 5000$ from Adison E. motter, Takashi
Nishikawa and Ying-Cheng Lai's work (Physical Review E
\textbf{66}, 065103(R) (2002)).

\begin{thebibliography}{99}
\bibitem{pexp} The RSF network considered here has totally $3Nz_1$ half-links. Pairs of half-links are chosed
and connectted total randomly, thus the probability of a pair of nodes with degree $k_{1}$ and $k_{2}$
connected directly should be equal to $p^{1}(k_{1,}k_{2})\approx \frac{k_1 k_2}{2Nz_1}$.
\end{thebibliography}
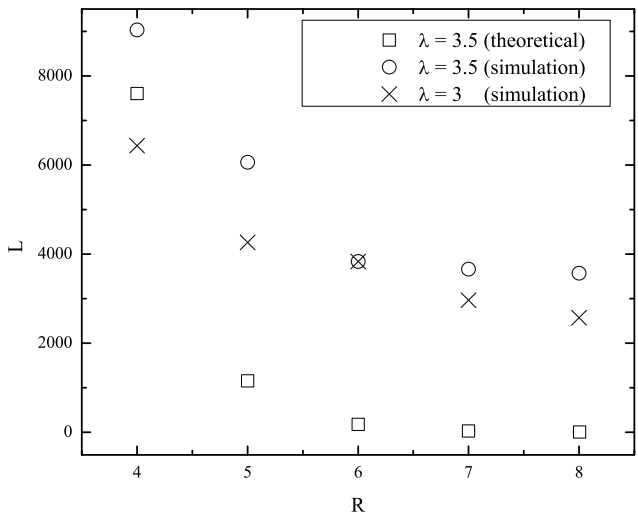
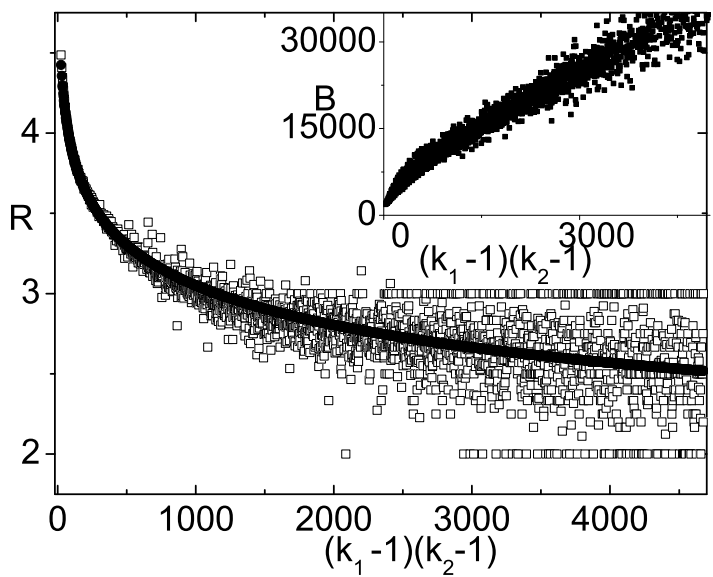
\end{document}